\documentclass[a4paper,fleqn,usenatbib]{mnras}

\usepackage{mathptmx}

\usepackage[T1]{fontenc}
\usepackage{ae,aecompl}

\usepackage{booktabs,tabularx} 

\usepackage{graphicx}	
\usepackage{amsmath}	
\usepackage{amssymb}	

\usepackage{multirow}
\usepackage{threeparttable}
\usepackage{caption}
\usepackage{dblfloatfix}
\usepackage{xcolor} 

\title[Hot Gas Flows in NGC\,3115]{Hot Gas Flows on Parsec Scale in the Low-Luminosity Active Galactic Nucleus NGC\,3115}

\author[Yao et al.]{
	Zhiyuan Yao,$^{1}$
	Zhaoming Gan$^{2,1}$\thanks{E-mail: z.gan@columbia.edu}
	\\
	$^{1}$Key Laboratory for Research in Galaxies and Cosmology, Shanghai Astronomical Observatory, Chinese Academy of Sciences,\\
  \ 80 Nandan Road, Shanghai 200030, China \\
  	$^{2}$Department of Astronomy, Columbia University, 550 W, 120th Street, New York, NY 10027, USA
}

\date{Accepted XXX. Received YYY; in original form ZZZ}

\pubyear{2018}

\begin{document}
\label{firstpage}
\pagerange{\pageref{firstpage}--\pageref{lastpage}}
\maketitle
	
\begin{abstract}
NGC 3115 is known as the low-luminosity active galactic nucleus which hosts the nearest ($z\sim0.002$) billion solar mass supermassive black hole ($\sim1.5\times10^9~M_\odot$). Its Bondi radius $r_\mathrm{B}$ ($\sim3\farcs6$) can be readily resolved with Chandra, which offers us an excellent opportunity to investigate the accretion flow onto a supermassive black hole. In this paper, we perform two-dimensional hydrodynamical numerical simulations, tailored for NGC 3115, on the mass flow across the Bondi radius. Our best fittings for the density and temperature agree well with the observations of the hot interstellar medium in the centre of NGC 3115. We find that the flow properties are solely determined by the local galaxy properties in the galaxy centre: (1) stellar winds (including supernova ejecta) supply the mass and energy sources for the accreting gas; (2) similar to the one-dimensional calculations, a stagnation radius $r_\mathrm{st}\sim0.1~r_\mathrm{B}$ is also found in the two-dimensional simulations, which divides the mass flow into an inflow-outflow structure; (3) the radiatively inefficient accretion flow theory applies well inside the stagnation radius, where the gravity is dominated by the supermassive black hole and the gas is supported by rotation; (4) beyond the stagnation radius, the stellar gravity dominates the spherical-like fluid dynamics and causes the transition from a steep density profile outside to a flat density profile inside the Bondi radius.

\end{abstract}
	
	


\section{Introduction}\label{sec:introduction}

It has reached an agreement that in the centre of every massive galaxy with a bulge harbours a supermassive black hole (BH) \citep{Kormendy:2013db}. Most of the BHs are not active with median Eddington ratio $L_\mathrm{bol}/L_\mathrm{Edd}\sim10^{-5}$ \citep{2009ApJ...699..626H}. The observed short active galactic nuclei (AGN) duty cycle \citep{Schawinski2015} can be readily explained by the suppression of large scale feedback caused by the AGN activities, leaving the galaxies low-luminosity for most time of their lives \citep{2011ApJ...737...26N, Yuan2018}.

Basically, LLAGNs are fed via two patterns. One is inflow from the gas travelling over galactic scales, which dominates in the galaxies abundant with gas such as spiral galaxies. The other is accretion through the stellar winds from nearby nuclear star clusters, which plays an important role in elliptical galaxies. In the latter case, collisions of ejected stellar winds in the feeding region around the Bondi radius produce hot gas with the temperature up to $10^7~\mathrm{K}$. Interactions with interstellar medium (ISM) do not cool down the gas, which will maintain the virial temperature $T_\mathrm{v}\sim0.3-1~\mathrm{keV}$ and radiate mostly in X-rays, leading to the following X-ray observations of \textit{Chandra}.

Nowadays, the exceptional X-ray spatial resolution of the \textit{Chandra} satellite allows for the study in detail of the gas flow within the Bondi radius in several nearby galaxies such as M31, M87, the Milky Way, and NGC\,3115. The Bondi radius $r_\mathrm{B}=2GM_\mathrm{BH}/c_{\mathrm{s},\infty}$, where $c_{\mathrm{s},\infty}$ is the adiabatic sound speed at infinity, is a characteristic radius to quantify the gravitational influence of the BH \citep{Bondi:1952zk}. Thus it is critical to resolve the Bondi radius so that the accretion flow could be observed in spite of the strong assumptions in the Bondi accretion theory. In this paper we focus on NGC\,3115, which has an accumulated exposure of $1$~Ms during the year $2012$ with \textit{Chandra} \citep{Wong:2014ko}. NGC\,3115 is an S0 lenticular galaxy, hosting the closest billion solar mass supermassive BH \citep{1992ApJ...393..559K,1996ApJ...459L..57K,1999MNRAS.303..495E} at a distance of about $d=9.7$~Mpc \citep{2001ApJ...546..681T}. The low temperature ($\sim 0.3$ keV) of the ambient gas implies a Bondi radius $r_\mathrm{B}=2\farcs4-4\farcs8$ that can be resolved readily with \textit{Chandra}. The AGN is only recently found in NGC\,3115 owing to radio observations \citep{2012AJ....144..160W}.  Deep studies of the centre reveal a very underluminous active galactic nucleus (with an upper limit of the Eddington ratio to be $L_\mathrm{X,AGN}/L_\mathrm{Edd} < 3.5\times10^{-10}(10^9M_{{\sun}} /M_\mathrm{BH})$ immersed in an old massive nuclear star cluster \citep{1992ApJ...393..559K, 1996ApJ...459L..57K,1999MNRAS.303..495E} , which is probably the feeding source of the AGN. 

Concerning the feeding process by nuclear star clusters, Sgr A* is a well investigated source due to its closest distance. Both simple one-dimensional calculations \citep{Quataert:zz,Shcherbakov:2010gs} and three-dimensional numerical simulations \citep{2008MNRAS.383..458C,2018arXiv180500474R} have shown that the observed stars could supply more than enough mass to explain the observed accretion rate and X-ray emissions. However, restricted to the resolution of \textit{Chandra}, we can not obtain the spatially resolved spectrum thus the radial profile of density or temperature across the Bondi radius for Sgr A*. With the X-ray luminosity alone, models are barely limited. NGC\,3115, one of very few sources with spatially resolved spectrum in the centre, provides us an excellent opportunity to investigate the supermassive BH feeding process. In particular, its density profile within the Bondi radius suggested by Chandra deviates from that predicted by the simulations.


The radial profiles of mass accretion rate and density of the hot gas in LLAGNs have been explored extensively by numerical simulations in the recent few years. The motivation is relevant to the fundamental physics behind this radiatively inefficient accretion flow (RIAF), which is found to be closely linked with LLAGNs \citep{1995ApJ...452..710N,1995Natur.374..623N} after the rediscovery of the self-similar solution for the advection-dominated accretion flow \citep{1994ApJ...428L..13N}. Subsequently with detailed study of its properties, RIAF models become established in the accretion literature. The leading model suggests that the mass accretion rate is a constant over the radius and the density profile follows $\rho\propto r^{-p}$ with $p=1.5$ \citep{1994ApJ...428L..13N}. \citet{Stone1999}, however, perform two-dimensional hydrodynamical simulations for the accretion of a rotating torus by adopting the ``$\alpha$'' description to transfer angular momentum. They find the mass accretion rate in the disk decreases with decreasing radius, and only the net accretion rate is constant over the radius just by being in a steady state. Correspondingly the density profile becomes flatter with $p<1.5$. Their results have been confirmed to be universal later by many numerical works with various setups\setcitestyle{notesep={; },round,aysep={},yysep={;}}\citep[see][and references therein]{Yuan2012a}. In order to explain this phenomenon of varying mass accretion rate, two competing models have been proposed: the ADIOS model \citep{1999MNRAS.303L...1B} and the convection-dominated accretion flow (CDAF) model (\citealt{2000ApJ...539..798N,2000ApJ...539..809Q}; see also the magnetically frustrated convection by  \citealt{2003ApJ...596L.207P,2011MNRAS.415.1228P}), which respectively emphasise the roles of two distinct processes in the RIAF: outflows and convection. \citet{Yuan2012b}, for the first time, show the existence of a global outflow by conducting and analysing the simulations with/without magnetic field. This is also suggested by observations \citep{2013Sci...341..981W,2016Natur.533..504C,2019MNRAS.483.5614M}, which highlight the presence of centrally driven winds in a large number of LLAGNs, particularly in our Milky Way. Moreover, the wind properties, such as mass flux and velocity, can be determined by theoretical studies \citep{Yuan2015,2018A&A...615A..35B}, and are proved to be important for the AGN feedback \citep{2010ApJ...722..642O,2017MNRAS.465.3291W,Yuan2018,2019ApJ...871..138B}.

Therefore, the  profiles of radial mass accretion rate and density gain attentions of theorists in the recent few years, and \citet{Yuan2012a} have provided a thorough discussion on this topic. Briefly speaking, almost all of the hot accretion flow simulations give radial density profiles $\rho \propto r^{-p}$ with $0.5<p<1$. The results depend weakly on the viscosity \citep{Bu2013} and magnetic field \citep{Yuan2012b}. \citet{Inayoshi2018} perform simulations on the accretion flow around the Bondi radius, and they also find the density slope $p<1$ across the Bondi radius regardless of viscosity and the angular momentum of the gas. NGC\,3115, however, suggests a steeper radial density profile with $p\sim1$ around the Bondi radius. Furthermore, two more nearby sources Sgr\,A* and M\,87 suggest $p\sim1$ in their central regions as well \citep{2019ApJ...871..126G,Russell:2015ec,Russell2018}. If the RIAF does exist inside the Bondi radius of these sources, something must miss in the simulations.

On the other hand, if the angular momentum of the gas is low, the density profile should return to the canonical Bondi solution. Yet its density profile strongly deviates from the observations \citep[Figure 8]{Wong:2014ko}. All these evidence suggests that stars might play a significant role in the accretion process, given the enclosed stellar mass approaches the BH mass around the Bondi radius. In other words, the gravitational potential contributed by stars as well as the stellar winds and supernovae as mass and energy sources might be important. Indeed, by considering the existence of stars, \citet{Shcherbakov:2014cz} perform one-dimensional calculations on the case of NGC\,3115 and can roughly fit the data. However, their models still have a variety of limitations owing to one-dimension as they have already pointed out at the end of their paper. First, they treat angular momentum transport as a posteriori without calculating explicitly. In fact, observations suggest that the rotational velocity of the stars at large radii can reach 0.8 times Keplerian velocity \citep{1992ApJ...393..559K}, indicating the large angular momentum of the hot gas ejected by the stars there. Hence the realistic gas flow may differ from one-dimensional calculations. Second, the gas flow in one-dimensional solutions is either inflow or outflow radially. With the polar direction in two-dimensional simulations, more complex patterns may occur, e.g., inflows around the equatorial plane and outflows along the polar axis. This pattern is more realistic and thus important for studying the accretion process in detail. Considering these reasons, we perform two-dimensional simulations with similar physics, attempting to form a more realistic picture and find out the role each physical process plays in the accretion episode, in particular on the radial density profile.


The structure of this paper is as follows. In Section~\ref{methods}, we first quantify properties of the nuclear star cluster in NGC\,3115 and explore the physical processes in the hot gas flow. Then we introduce the hydrodynamical equations with the initial and boundary conditions. In Section~\ref{sec:results}, we present our best-fitting results compared with the observations and explore each physical process in our model. In Section~\ref{sec:conclusions}, we summarise our results.

\section{Numerical Methods}\label{methods}

In this section, we present the framework of the model. Since the simulation region is around the Bondi radius, stellar gravitational potential can not be neglected. In fact, the gravitational contribution from stellar mass is roughly comparable to that from the supermassive BH at the Bondi radius. Stellar winds produced by the stars during the asymptotic giant branch (AGB) phase are the only mass source of accretion, yet they are sufficient to explain the X-ray flux of the observations. At the end of the stellar evolution, stars inject most of their mass and energy back to the ISM as supernova feedback. Type Ia supernovae occur more frequently than other kinds in the nucleus of S0 galaxies (like NGC\,3115) due to the large stellar population age. Apart from the stellar physics, cooling is considered in the simulations motivated by the drop of fitting temperature with decreasing radius in \citet{Wong:2014ko} if using a single temperature model. Moreover, thermal conduction is also incorporated since it should play an important role in collisionless plasma such as the RIAF with low accretion rate \citep{Johnson:2007ju}. \cite{Shcherbakov:2014cz} highlights its effect but we think it is overestimated as discussed below. The details of our physics are given as follows.

\subsection{Properties of the Nuclear Star Cluster} 
Nuclear star clusters are ubiquitous in galaxies with central supermassive BHs. Due to the close distance to the centre, they have great effects on the BH feeding naturally. Therefore, the properties of the nuclear star clusters, such as the stellar mass and gas injection with mass, energy, and angular momentum are required to be investigated at the beginning. Fortunately, the NGC\,3115 nuclear star cluster was extensively observed in the optical band in search of a supermassive BH with both ground-based instruments \citep{1992ApJ...393..559K} and the \textit{Hubble} Space Telescope \citep{1996ApJ...459L..57K,1999MNRAS.303..495E}. Readers are also referred to \cite{Shcherbakov:2014cz} for detailed descriptions of the nuclear star cluster of NGC\,3115.

\subsubsection{Enclosed Mass}
The radial profile of the enclosed stellar mass is able to be inferred from surface brightness profile \citep{1992ApJ...393..559K} by adopting the assumption of a constant mass-to-light ratio $\Upsilon_*=M_*/L_\mathrm{V}=4.0$ (in solar unit) ratio at any radius. The mass-to-light ratio $\Upsilon_*$ is derived from the ratio of the integrated mass to the luminosity interior to a radius $r$ as a function of $r$, which is almost constant outside the Bondi radius \citep{1992ApJ...393..559K}. The derived enclosed stellar mass is comparable to the BH mass at the Bondi radius.

\subsubsection{Mass Injection}\label{mass_inj}
It is commonly believed that in the nucleus of the early-type galaxy with low mass accretion rate, like NGC~3115, stellar winds play a dominant role in the accretion process. The one-dimensional study has already shown that stellar winds supply more than enough mass to explain the emissions \citep{Shcherbakov:2014cz}. Moreover, compared with Sgr~A* where only 30 or so Wolf-Rayet stars contribute substantially \citep{2006ApJ...643.1011P}, two orders of magnitude more massive stars exist in the central parsec of NGC\,3115, which is derived from the enclosed stellar mass assuming Kroupa initial mass function \citep{2001MNRAS.322..231K}. Therefore, we assume stellar winds to be the only mass source in the model. Its mass loss rate is quantified to be proportional to the stellar mass and to the stellar luminosity \citep{1991ApJ...376..380C,2001A&A...376...85J,2002ApJ...571..272A}. Either is equivalent on account of the assumption of a constant ratio of stellar mass to luminosity. Here we adopt the correlation with the $V$-band luminosity
\begin{equation}\label{eq:mdot}
\dot{M}_* \approx 3\times10^{-11}\left(\frac{L_\mathrm{V}}{L_{{\sun}\mathrm{,V}}}\right){M_{\sun}\mathrm{yr}^{-1}}
   = 3\times10^{-11}\Upsilon_*^{-1}{M_*\mathrm{yr}^{-1}}
\end{equation} 
for an old stellar population \citep{1976ApJ...204..365F,1993ApJ...416...26P}. We normalise this mass loss rate by a radius-independent parameter $f_\mathrm{q}$ on account of the uncertainties in estimating the mass loss rate.

\subsubsection{Energy Injection}
The energy injection in our model is twofold: stellar winds and Type Ia supernova feedback. Although the wind characteristic velocity is under $50~\mathrm{km~s}^{-1}$\citep{1982ApJ...252..616K,2009PASA...26..365M,2010A&A...510A..14L,2011ApJ...734...48L} for AGB stars that contribute stellar winds most in old star clusters, the wind velocity with respect to the ISM can be much higher since the mass-shedding stars in the NGC\,3115 nucleus could be of the velocity dispersion around $300~\mathrm{km~s}^{-1}$ in a combined gravitational field of the supermassive BH and the enclosed stellar mass. Therefore, we ignore the wind characteristic velocity and arrive at the wind velocity due to the motion of stars
\begin{equation}
v_\mathrm{w,st}=\sqrt{{G\left(M_\mathrm{enc}+M_\mathrm{BH}\right)}/r}.
\end{equation} 
This velocity is consistent with the measurements of dispersive velocity in the observations \citep{1992ApJ...393..559K}. 

Type Ia supernovae should contribute to the total energy injection as well. \cite{Shcherbakov:2014cz} estimate the specific energy injection rate by supernovae and find it larger than by stellar winds with a typical velocity $\sim300~\mathrm{km~s}^{-1}$. Following their assumption of effective supernova wind velocity, we adopt the equivalent velocity $v_\mathrm{w,sn}$ that the wind has if its energy equals that injected by supernovae. The total energy injection from stellar winds and Type Ia supernovae is thus described by
\begin{equation}
\label{eq:vw}
v_\mathrm{w}=\sqrt{v^2_\mathrm{w,st}+v^2_\mathrm{w,SN}}.
\end{equation}

\subsubsection{Angular Momentum Injection}
Based on the observations\citep{1992ApJ...393..559K}, the NGC\,3115 nuclear star cluster possesses a non-zero mean rotation. The rotational profile can be described as:
\begin{equation}\label{eq:vphi}
v_{\phi}=257\left(\frac{w}{20\arcsec}\right)^{0.287}\mathrm{km~s}^{-1},
\end{equation} 
where $w=r~\mathrm{sin}\theta$ is the cylindrical radius. The rotational velocity can be as high as 0.8 times Keplerian velocity at the outer boundary of our simulation region, yet quickly approaches zero at a small $w$.

\subsection{Physical Effects on the Gas Flow}
\subsubsection{Gravitational Pull by the Enclosed Stellar Mass}
According to observations, the mass of central supermassive BH lies in the range $(1-2)\times10^9~M_{\sun}$ \citep{1996ApJ...459L..57K,2011ApJ...736L..23W,Wong:2014ko}. In our simulations we choose $M_\mathrm{BH}=1.5\times10^9~M_{\sun}$. The Bondi radius can be estimated by $r_\mathrm{B}\equiv 2GM_\mathrm{BH}/c^2_{\mathrm{s},\infty}=168~\mathrm{pc}=3\farcs6$ (the sound speed is $c_{\mathrm{s},\infty}\simeq275~\mathrm{km/s}$). Even though this giant monster has a crucial influence on its neighbours, the gravity of the enclosed stellar mass, however, is comparable to that of the supermassive BH at the Bondi radius, indicating the significance of the gravitational pull by the stars. The total gravitational potential can be written as
\begin{equation}
	\psi=-\frac{G\left(M_\mathrm{enc}+M_\mathrm{BH}\right)}{r}.
\end{equation}

\subsubsection{Radiative Cooling}
\textit{Chandra} observes a temperature drop inside the Bondi radius, suggesting a crucial role of cooling in the accretion process. We employ the CIE cooling curve from \citet{1993ApJS...88..253S} by ignoring the effects of clumping and spatial inhomogeneity and assume a metallicity of $Z/Z_{\sun}=1$. The cooling power per unit volume reads
\begin{equation}
\mathcal{C}=n_\mathrm{e} n_\mathrm{i} \Lambda(T,Z).
\end{equation}

\subsubsection{Thermal Conduction}
Conduction and convection are of capacity to transfer energy from the inner flow to the outer flow. Convection appears to be important in collisional flows \citep{2000ApJ...539..798N,2000ApJ...539..809Q} while conduction begins to dominate in collisionless flows \citep{Shcherbakov:2010gs}. The key physical quantity is the mean free path
\begin{equation}
\lambda \approx0.23(\frac{T}{10^7~\mathrm{K}})^2(\frac{n_\mathrm{e}}{1~\mathrm{cm}^{-3}})^{-1}~\mathrm{pc}.
\end{equation}
\citet{Wong:2014ko} provides the profiles of temperature and density of the ISM in the nuclear region of NGC\,3115 by fitting observational data, from which the mean free path can be calculated and the flow is marginally collisional in this region. Following \citet{Shcherbakov:2014cz}, we adopt the unsaturated conduction with the heat flux
\begin{equation}
{\bf Q}_\mathrm{cond} = -\kappa  \nabla  T_\mathrm{e},
\end{equation}
where the thermal conductivity $\kappa$ satisfies \citep{1977ApJ...211..135C}
\begin{align}
\kappa=1.31n_\mathrm{e}k\lambda(kT_\mathrm{e}/m_\mathrm{e})^{1/2}.
\end{align}
Since $\lambda/r$ varies between $0.005-0.03$ in the region with observational data ($0\farcs5-17\arcsec$), we choose $\lambda=0.01/1.31~r$ across the simulation region and arrive at 
\begin{equation}\label{eq:conduction}
\kappa=0.01n_\mathrm{e}kr(kT_\mathrm{e}/m_\mathrm{e})^{1/2}.
\end{equation}
Note that the conductivity is approximately an order of magnitude smaller than that in \citet{Shcherbakov:2014cz}, since they use $0.1$ instead of $0.01$ in Equation~(\ref{eq:conduction}) without giving specific explanations.

\subsection{Hydrodynamics}
\subsubsection{Equations}
We solve the following time--dependent Eulerian equations of hydrodynamics with stellar source terms, thermal conduction and radiative cooling:
\begin{align} \begin{split}
	\frac{\partial \rho}{\partial t} + \nabla\cdot(\rho{\bf v}) &= f_\mathrm{q}\dot{\rho}_*,\\
	\frac{\partial \rho{\bf v}}{\partial t}+\nabla\cdot(\rho{\bf v v}+P{\bf I}) &=- \rho\nabla\psi+ f_\mathrm{q}\dot{\rho}_*{\bf v}_*+ \nabla\cdot\bf T,\\
   	\frac{\partial E}{\partial t} + \nabla\cdot((E+P){\bf v}) &=
   	 -\mathcal{C} - \nabla \cdot {\bf Q}_\mathrm{cond}\\
	&+\frac{1}{2}f_\mathrm{q}\dot{\rho}_* \left(v_\mathrm{w} ^2+{\bf v}_*^2\right)+{\bf T}^2/\mu, 
\end{split} \end{align}
where $\rho$ and $P$ are the gas density and pressure, respectively. $E=1/2\rho v^2+P/(\gamma-1)$ is the total energy per unit volume, and $\gamma=1.5$ is the adiabatic index of the gas. We choose $\gamma<5/3$, otherwise the sonic point for the initial Bondi flow would be smaller than the inner boundary, thus producing spurious reflection of waves on the inner boundary due to subsonic flow \citep{1991JCoPh..94....1G}. We have verified the exact value of $\gamma$ do not affect our main results.
$\psi$ is the gravitational potential of the stars and the central supermassive BH. The self-gravity of the accretion flow is neglected. $\dot{\rho}_*$ is the stellar mass loss rate $\dot{M}_*$ (Equation (\ref{eq:mdot})) per unit volume and satisfies $d\dot{M}_*/dr = 4\pi r^2\dot{\rho}_*$. ${\bf v}_*$ is the velocity of the mass source with only a non-zero azimuthal component $v_{\phi}$ (Equation (\ref{eq:vphi})). Apart from the rotational velocity $v_{\phi}$, stellar winds possess dispersive velocity $v_\mathrm{w}$ as well. The wind velocity $v_\mathrm{w}$ is a combination of the dispersive velocity from stars $v_\mathrm{w,st}$ and the supernova effective velocity $v_\mathrm{w,sn}$ as given in Equation (\ref{eq:vw}). The rotational velocity is significant in terms of the energy and momentum at large radii while the dispersive velocity contributes to the energy at small radii. $\mathcal{C}$ and ${\bf Q}_\mathrm{cond}$ represent cooling and thermal conduction, respectively.

Considering the lack of magnetic field in our simulations, we employ the ``$\alpha-$viscosity'' scheme to transfer angular momentum. Here, $\bf T$ is anomalous stress tensor. Following \citet{Stone1999}, we assume only azimuthal components of the stress are non-zero:
\begin{equation}
	T_{r\phi}=\mu r\frac{\partial}{\partial r}\left(\frac{v_\phi}{r}\right), 
\end{equation}
\begin{equation}
	T_{\theta\phi}=\frac{\mu \sin\theta}{r}\frac{\partial}{\partial\theta}\left(\frac{v_\phi}{\sin\theta}\right).
\end{equation}
 The viscosity coefficient $\mu=\nu\rho$ determines the magnitude of the stress and $\nu$ is kinematic viscosity. We adopt the form $\nu=\alpha \sqrt{\mathrm{G}M_\mathrm{BH}r}$ (\citet{Stone1999}, their model~K) and $\alpha=0.01$. 

\subsubsection{Initial Setup and Boundary Conditions}
We use the Bondi solution as the initial condition, which has a temperature of $0.3$~keV and extremely low density at ``infinity''. We also test the uniform initial condition, and find no difference to our results. This is because the injection timescale is an order of magnitude smaller than the evolutionary time as shown in Figure \ref{fig:timescale}. The evolutionary time is chosen to allow the gas at the outermost radius with the circularisation radius $r_\mathrm{cir,out}=r_\mathrm{out}v_\phi(r_\mathrm{out})/v_\mathrm{K}(r_\mathrm{cir,out})\sim6\farcs7$ flowing into the centre, which is $r_\mathrm{cir,out}/\alpha v_\mathrm{K}(r_\mathrm{cir,out})\sim200$~Myr, if the RIAF theory applies in the simulation region. We perform two-dimensional hydrodynamics simulations with ZEUS-MP/2 \citep{2006ApJS..165..188H} in spherical coordinates ($r$, $\theta$, $\phi$), i.e., all quantities are assumed to be axisymmetric with $90\times30\times1$ grid cells. We perform a simulation test with higher resolution of $240\times84\times1$ for our best fitting model and the results remain unchanged. The simulation domain extends from $r_\mathrm{min}=0\farcs004$ ($\sim10^{-3}~r_\mathrm{B}$) to $r_\mathrm{max}=17\arcsec$ ($\sim5~r_\mathrm{B}$) with $\theta$ from 0.1 to 3.04, $\phi$ from 0 to $2\pi$. The inner boundary is chosen to be smaller than the sonic radius for the initial Bondi solution to limit the spurious reflection of waves on the inner boundary due to subsonic flow \citep{1991JCoPh..94....1G}, while the outermost grid radius is limited by the spatial range of observations on the enclosed stellar mass in \cite{1992ApJ...393..559K}. We carve the polar region from the simulations to avoid the singularity at the pole. A logarithmic mesh is adopted in the radial direction to gain high resolution in the inner computational region, while in the polar direction the mesh is divided homogeneously. We use the standard ``outflow boundary condition'' in the ZEUS code both on the inner/outer radial boundary and the polar boundary (see \citet{Stone:1992un} for more details).

\section{Results and Discussions}\label{sec:results}

In this section, we first discuss the timescales of each physical process to form an instructive picture and then introduce our simulation results in detail. Next, we compare our results with observations and find a remarkable consistency. To explore the roles of various physical processes on the density profile, we include single physics in turn in order to study its effect. Finally, we compare our two-dimensional simulation results with one-dimensional calculations.

\subsection{Timescales of physical processes}
\label{sec:timescale}
\begin{figure}
\centering
\includegraphics[width=0.5\textwidth]{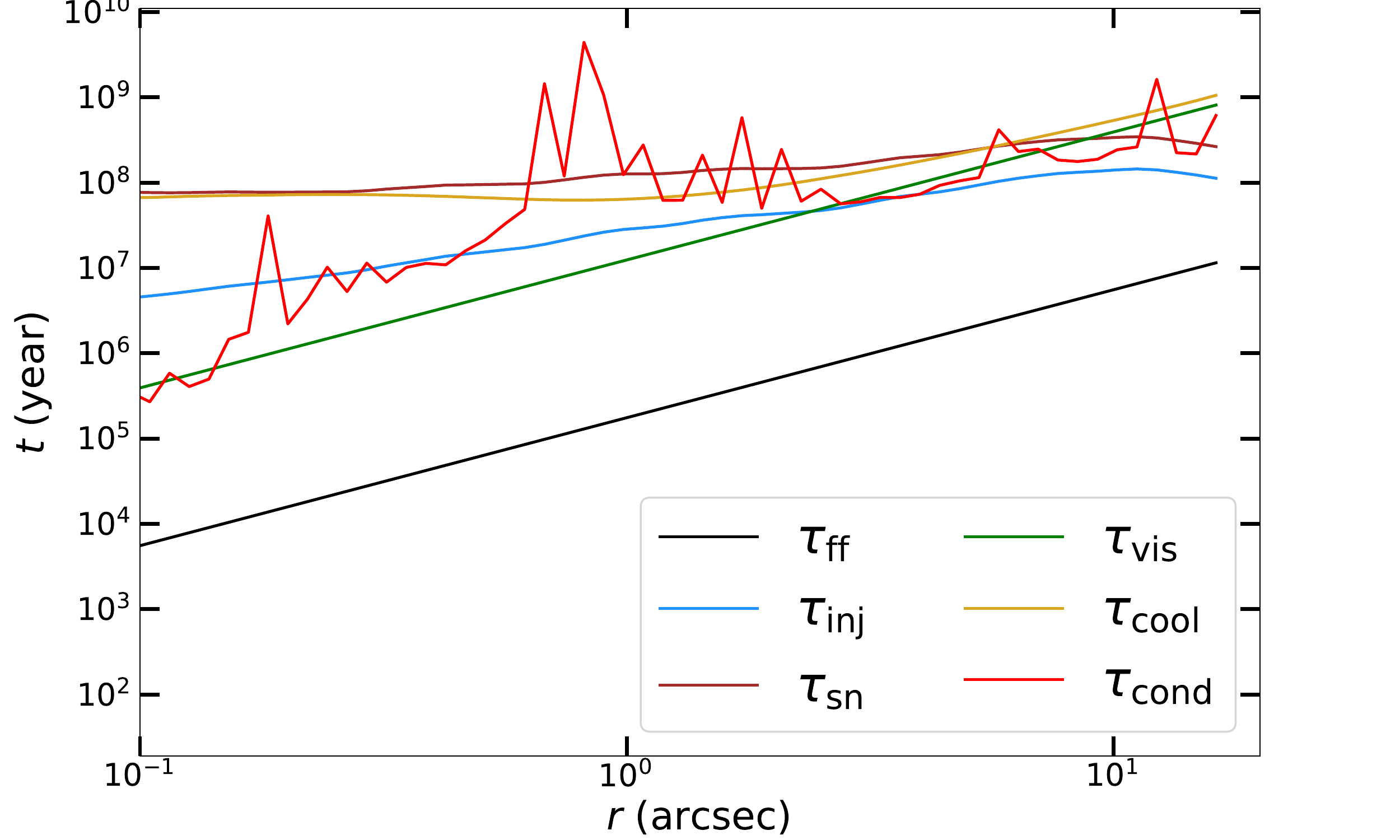}
\caption{
Timescales as functions of the radius: the free-fall timescale $\tau_\mathrm{ff}$ (black line), the mass injection timescale $\tau_\mathrm{inj}$ (blue line), the supernova timescale $\tau_\mathrm{sn}$ (brown line), the viscous timescale $\tau_\mathrm{vis}$ (green line), the cooling timescale $\tau_\mathrm{cool}$ (golden line), and the conductive timescale $\tau_\mathrm{cond}$ (red line). These timescales are averaged over the whole angles and time between 190-200~Myr.
}
\label{fig:timescale}
\end{figure}

It is worthwhile to begin our work with timescales since they are instructive and intuitive. Here we discuss several timescales corresponding to the physical processes included in our simulations, such as radiative cooling, unsaturated thermal conduction, mass and energy injection from stellar wind and supernovae. The cooling timescale is described by
\begin{equation}
	\tau_\mathrm{cool}=\frac{E}{\mathcal{C}}=\frac{E}{n_\mathrm{e} n_\mathrm{i} \Lambda(T,Z)},
\end{equation}
and the conductive timescale is given by
\begin{equation}
	\tau_\mathrm{cond} = \frac{E}{\nabla \cdot {\bf Q}_\mathrm{cond}}.
\end{equation}
The mass injection timescale by stellar mass is
\begin{equation}
	\tau_\mathrm{inj}=\frac{\rho}{f_\mathrm{q}\dot{\rho}_*}, 
\end{equation}
and the energy injection timescale by supernovae follows
\begin{equation}
	\tau_\mathrm{sn}=\frac{2E}{f_\mathrm{q}\dot{\rho}_*v_\mathrm{w,sn}^2}.
\end{equation}
Finally, we also show the free-fall timescale and viscous timescale for comparison
\begin{align}
	&\tau_\mathrm{ff}=\sqrt{\frac{2r^3}{GM_\mathrm{BH}}}, \\
	&\tau_\mathrm{vis}=\frac{r^2}{\nu}=\frac{r^2}{\alpha\sqrt{GM_\mathrm{BH}r}}.
\end{align}
Here all these timescales are averaged over angles. The radial-dependent and time-averaged timescales between 190-200~Myr are shown in Figure \ref{fig:timescale} with $f_\mathrm{q}=0.02$ and $v_\mathrm{w,sn}=350~\mathrm{km~s}^{-1}$, which are the parameters giving the best density and temperature fittings to the observations. 

\begin{figure*}
\centering
\includegraphics[width=\textwidth]{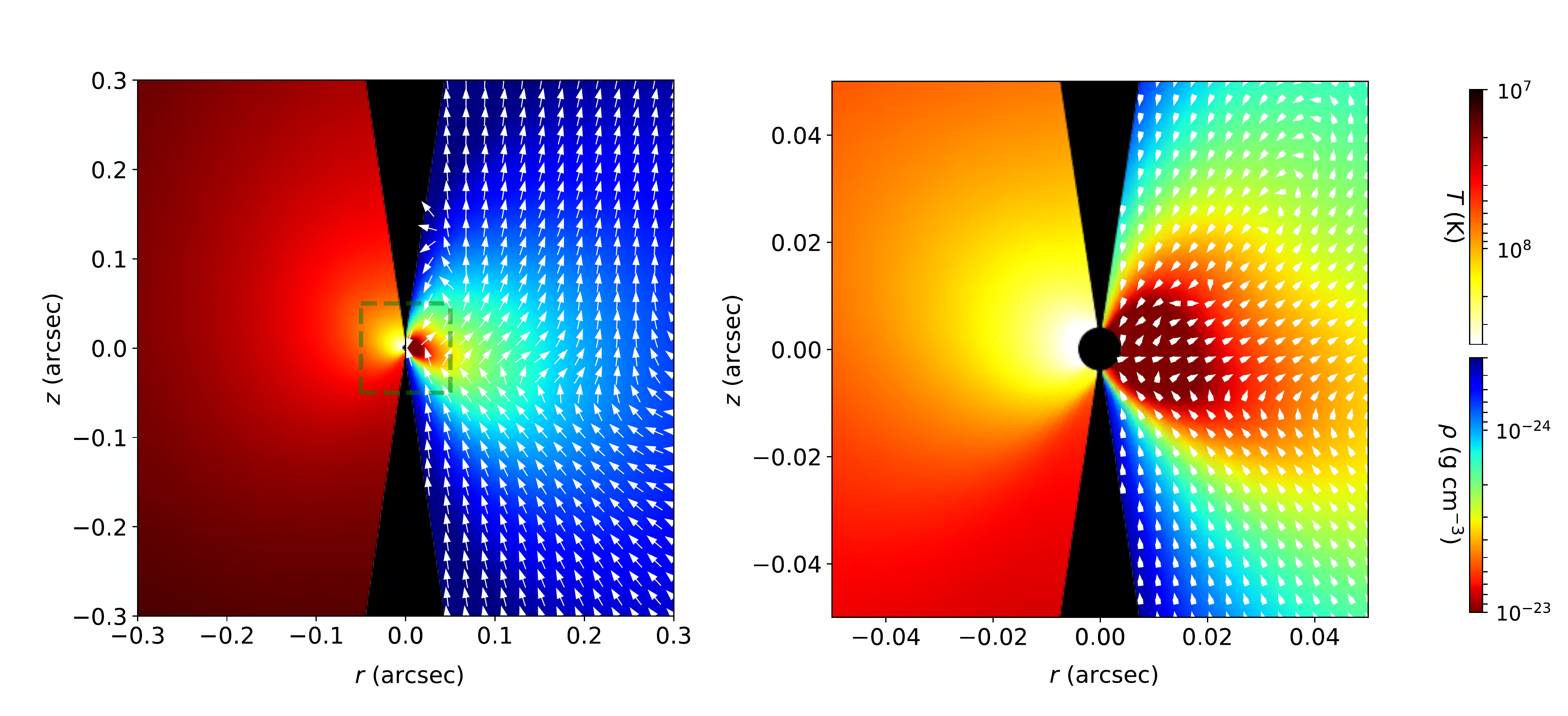}
\caption{Pseudocolors of temperature and density in the inner region with velocity over-plotted at 200~Myr. The right panel is a zoom-in plot of the central region marked by green dashed line in the left panel.
}
\label{fig:DensTemp}
\end{figure*}

In this figure , we can see that the cooling timescale shown as a golden line is over 100 times larger than the free-fall timescale and 10 times larger than the viscous timescale at almost any radius, indicating the inefficient cooling for the growth of thermal instability \citep{2012MNRAS.420.3174S,Gaspari:2013gz}. Meanwhile, the conductive timescale represented by a red line is generally smaller than the cooling timescale, suppressing the growth of instability in the local regions. In particular, the conductive timescale suddenly drops drastically inside 1\arcsec, because the unsaturated conductive heat flux is extremely sensitive to the temperature ($\propto T^{7/2}$, \citet{1962pfig.book.....S}). Yet according to the argument in \citet{1977ApJ...211..135C}, this is unphysical since the mean free path $\lambda$ is larger or comparable to the temperature scale height $T/\left| \nabla T\right|$ when $r\lesssim0\farcs01$, in the case of which the saturated conductive heat flux should be adopted instead. Considering the uncertainties of thermal conduction caused by magnetic field, two-temperature plasmas, and vertical temperature gradient, we still use the unsaturated form of the thermal conductivity in this region for simplicity, but we claim that our primary conclusions of this paper are not influenced by this strong conduction at small radii. The injection timescale denoted by a blue line is lower than evolutionary time across the whole radius, suggesting the weak dependence on the initial condition of the gas properties. The supernova timescale shown as a brown line is almost constant with radius. Since other timescales become larger with increasing radius, the supernova feedback should play a more important role at large radii. 

\subsection{Simulation results}
\label{sec:simulation}

\begin{figure}
\centering
\includegraphics[width=0.5\textwidth]{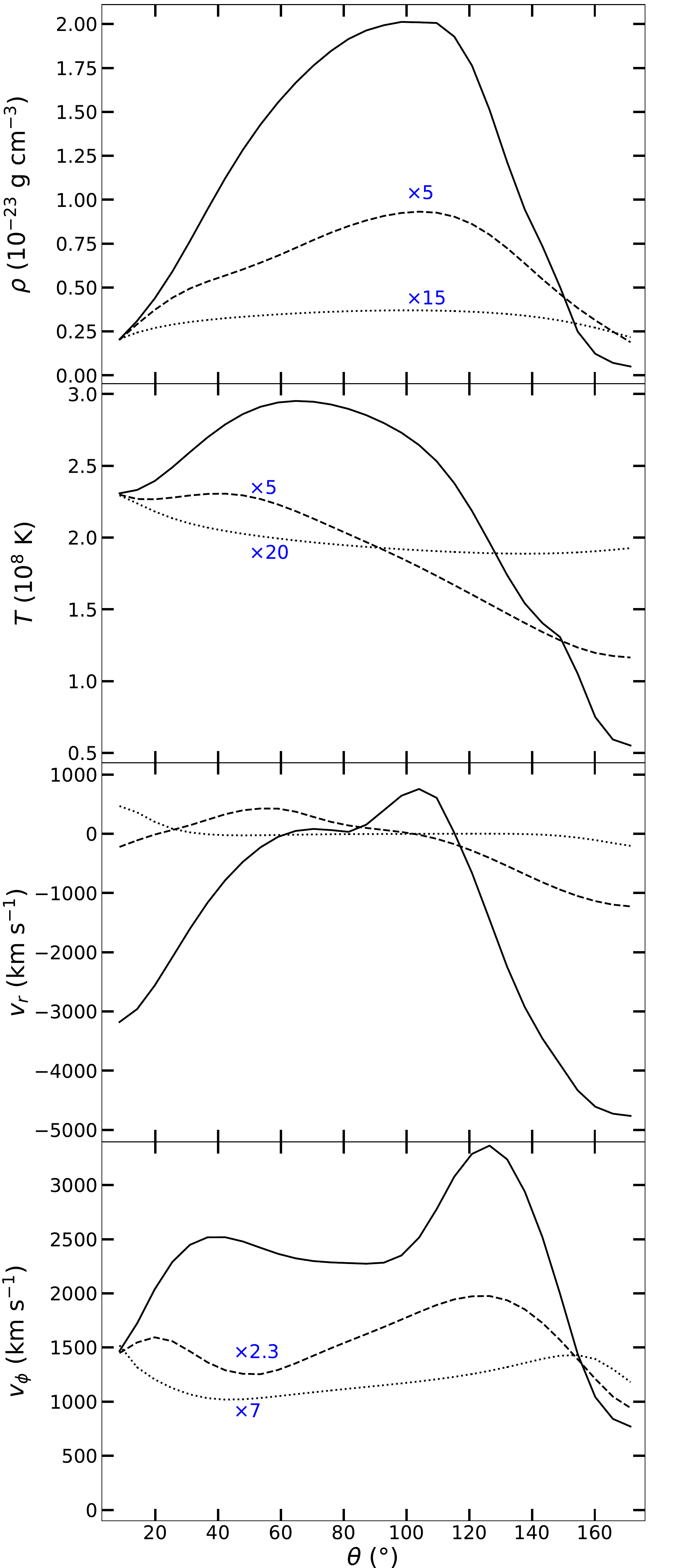}
\caption{Angular profiles of the density (the top panel), temperature (the second panel), radial velocity (the third panel), and rotational velocity (the bottom panel) at 200~Myr and radial positions of $r=0\farcs01$ (solid line), $r=0\farcs1$ (dashed line), and $r=1\arcsec$ (dotted line). Several lines are multiplied a given number for the convenience of comparison.
}
\label{fig:Angle}
\end{figure}

\begin{figure}
\centering
\includegraphics[width=0.5\textwidth]{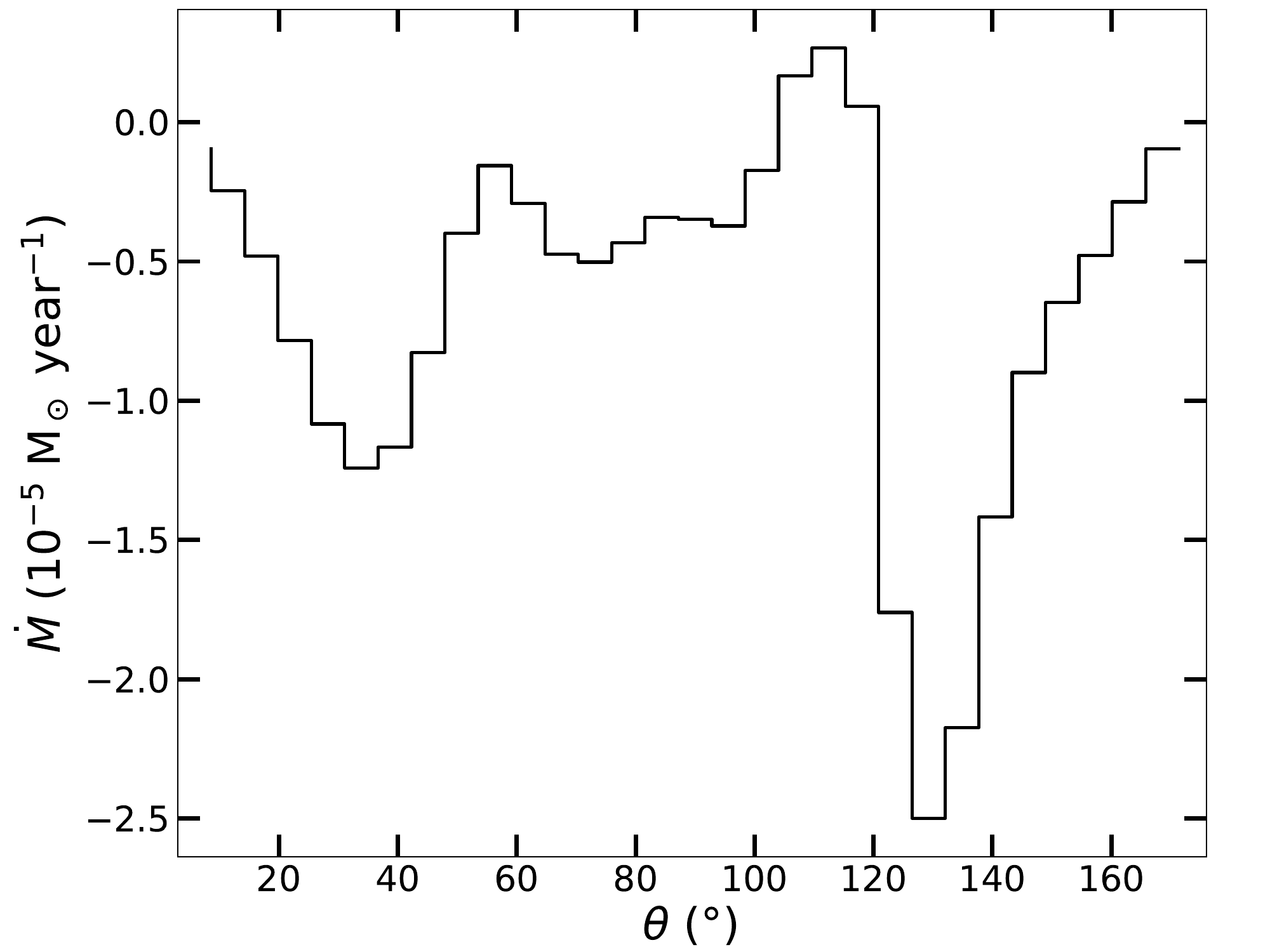}
\caption{Angular profiles of the mass accretion rate. The polar angles are divided by 30 bins, and the mass accretion rate is calculated through each bin at the inner boundary at 200~Myr.  As shown in this plot, the mass accretion rate is dominated by the accretion at the surface of the thick disk.
}
\label{fig:mass_accretion}
\end{figure}

\begin{figure}
\centering
\includegraphics[width=0.5\textwidth]{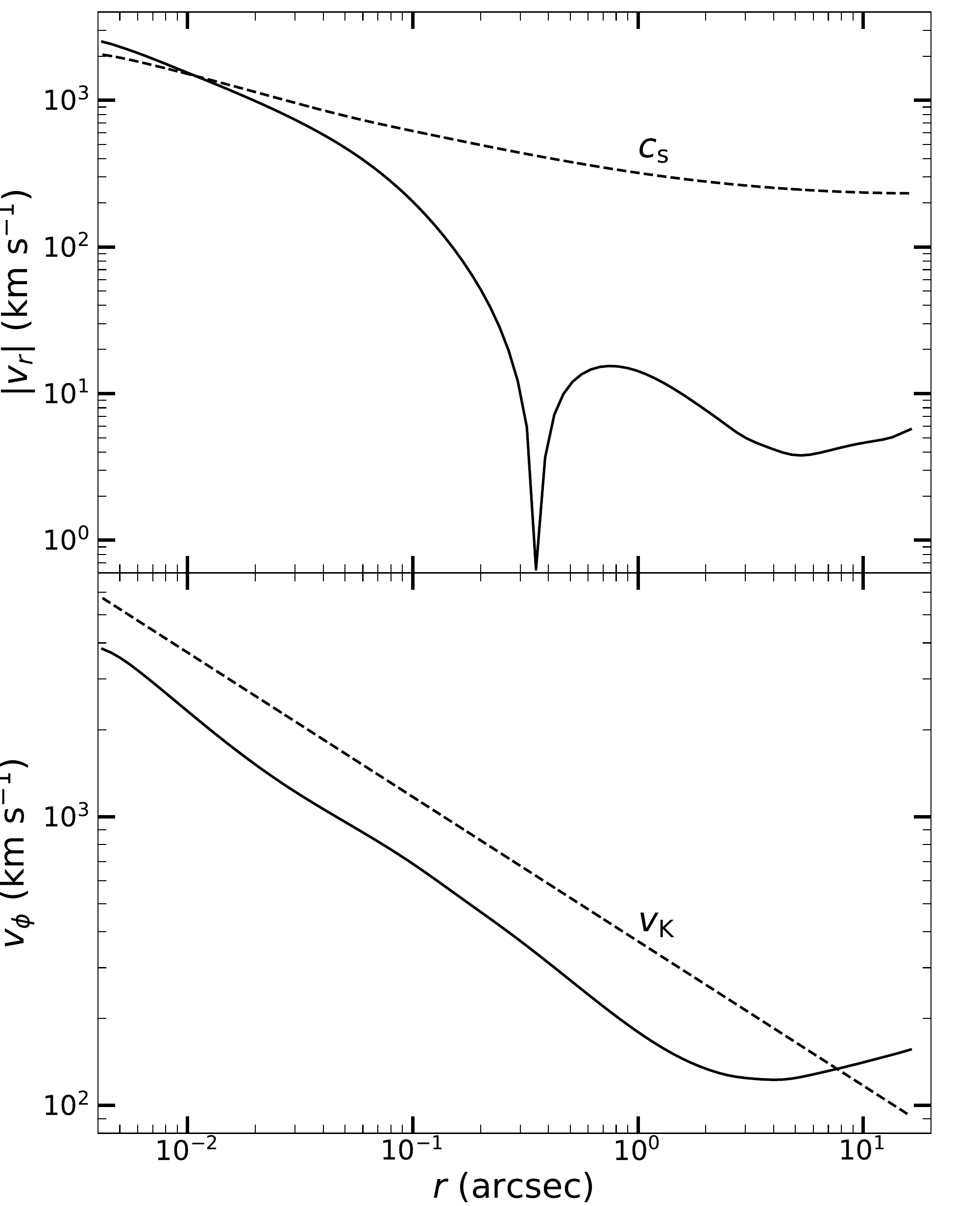}
\caption{Radial velocity and rotational velocity profiles averaged over the whole angles at 200~Myr. A stagnation radius $r_\mathrm{st}$ can be found at $\sim0.1~r_\mathrm{B}$ in the top panel. Inflows occur inside $r_\mathrm{st}$ while outflows dominate outside $r_\mathrm{st}$. The dashed line represents the sound speed in the top panel, and denotes the Keplerian velocity in the bottom panel.
}
\label{fig:Vel}
\end{figure}

Before fitting our simulation data to the observations, we present a visual impression of our simulations. Figure \ref{fig:DensTemp} shows the density and temperature pseudocolors with velocity over-plotted at 200~Myr. The right panel is a zoom-in plot of the central region marked by green dashed line in the left panel. The flow has entered into the steady state, which is judged by two criteria: the mass accretion rate at the inner boundary reaches constant over time and the net mass accretion rate does not vary with radius. Overall, the density and temperature are fairly smooth and spherically symmetric, except in the innermost region as shown in the top two panels of Figure~\ref{fig:Angle}. Lines at large radii are multiplied a given number for the convenience of comparison. At $r=0\farcs01$ the gas density in the equatorial plane is an order of magnitude higher than that at the poles, and the scale height (characterised by the density decreasing by a factor of $e$) is $\sim0\farcs017$. Hence the opening angle of this thick disk is $\sim120\degr$. The difference of temperature is smaller than that of density, and there is no symmetry with respect to the equatorial plane. At the first 1 Myr the system is fairly symmetric with the outflow propagates along the midplane. But soon the outflow turns over in the central region due to some asymmetric, unstable modes that breaks the symmetry. Above the midplane the gas in the polar region flows in directly, bringing the low-temperature gas outside into the central region and causing the lower temperature. The high velocity of inflow also decreases the density. Below the midplane the flow in the polar region is divided into two parts: inside 0.1\arcsec the gas flows inward due to strong BH gravity; outside 0.1\arcsec the gas is mixed with the outflow as it becomes much stronger and the opening angle of the outflow also increases. In the case without thermal conduction, this asymmetry fluctuates and almost disappears if we do the time average. With conduction, however, this asymmetry persists in time as the simulation enters the steady state due to the suppression of the random turbulence.

In terms of the gas kinematics, the angular profile of radial velocity in the third row of Figure~\ref{fig:Angle} indicates the accretion flow in the central polar regions. This can be seen directly in the right panel of Figure~\ref{fig:DensTemp}. The reason is relevant to the low angular momentum of the stellar winds in the polar region with small cylindrical radii, which is confirmed by the bottom panel of Figure~\ref{fig:Angle} that shows the angular profile of rotational velocity. Without centrifugal force to balance the gravity, gas is accreted directly at the poles. This ``funnel'' accretion scenario is also found in other simulations with spherical-like distribution of rotationally accreting material \citep{Proga2003,2018arXiv180500474R}. Note that they do not include viscosity in their simulations, indicating that this accretion pattern is not concerned with viscosity. Figure~\ref{fig:mass_accretion} shows the angular profile of mass accretion rate at the inner boundary at 200~Myr. It is interesting to find that the mass accretion rate is dominated by the accretion at the surface of the thick disk given the opening angle of the disk $\sim120\degr$. Although the velocity of the accreting gas is quite high near the pole, its low density restricts the mass accretion rate. Within the disk, the gas flows outward driven primarily by azimuthal stresses associated with the shear viscosity \citep{2013ApJ...767..105L}.

The angle-averaged radial velocity profile shows a similar inflow-outflow structure to the one-dimensional calculations \citep{Shcherbakov:2014cz,Generozov2015} with the stagnation radius $r_\mathrm{st}\sim0.1~r_\mathrm{B}$ (the top panel of Figure \ref{fig:Vel}). Here the stagnation radius is defined by a radius where volume-averaged radial velocity of the gas over the polar direction equals zero. In comparison with one-dimensional calculations, the gas flow can still move inward or outward at the stagnation radius owing to the additional angular structure. This is more plausible in the realistic environment since gas at large radii is required to feed the central BH. The bottom panel of Figure \ref{fig:Vel} shows that inside the stagnation radius, the rotational velocity is a half of the Keplerian velocity over the radius and the radial velocity tends to be supersonic and comparable to the rotational velocity as the radius decreases. Therefore, the flow is advection-dominated and radiatively inefficient (shown as timescales in Figure~\ref{fig:timescale}) during the accretion process, conforming to the standard RIAF theory.


\subsection{Fitting observational data}
\label{sec:BestFitting}

To compare our simulation results of density and temperature with observations strictly, we should first extend our two-dimensional simulations to three-dimension by rotating around the axis, then conduct the projection along the line of sight to acquire a circular region, which is divided into several circular rings same as the observation and is calculated the simulated spectrum of the X-ray emission in each ring, and finally fit the spectra by using the X-ray Spectral Fitting Package (XSPEC) to obtain the projected density and temperature profiles. However, since these procedures are cumbersome and \citet{Wong:2014ko} have already provided the deprojected density profile, we simply compare our simulated density with observations directly and choose surface brightness instead of temperature to be the second and sufficient observational constraint.

\begin{figure}
\centering
\includegraphics[width=0.5\textwidth]{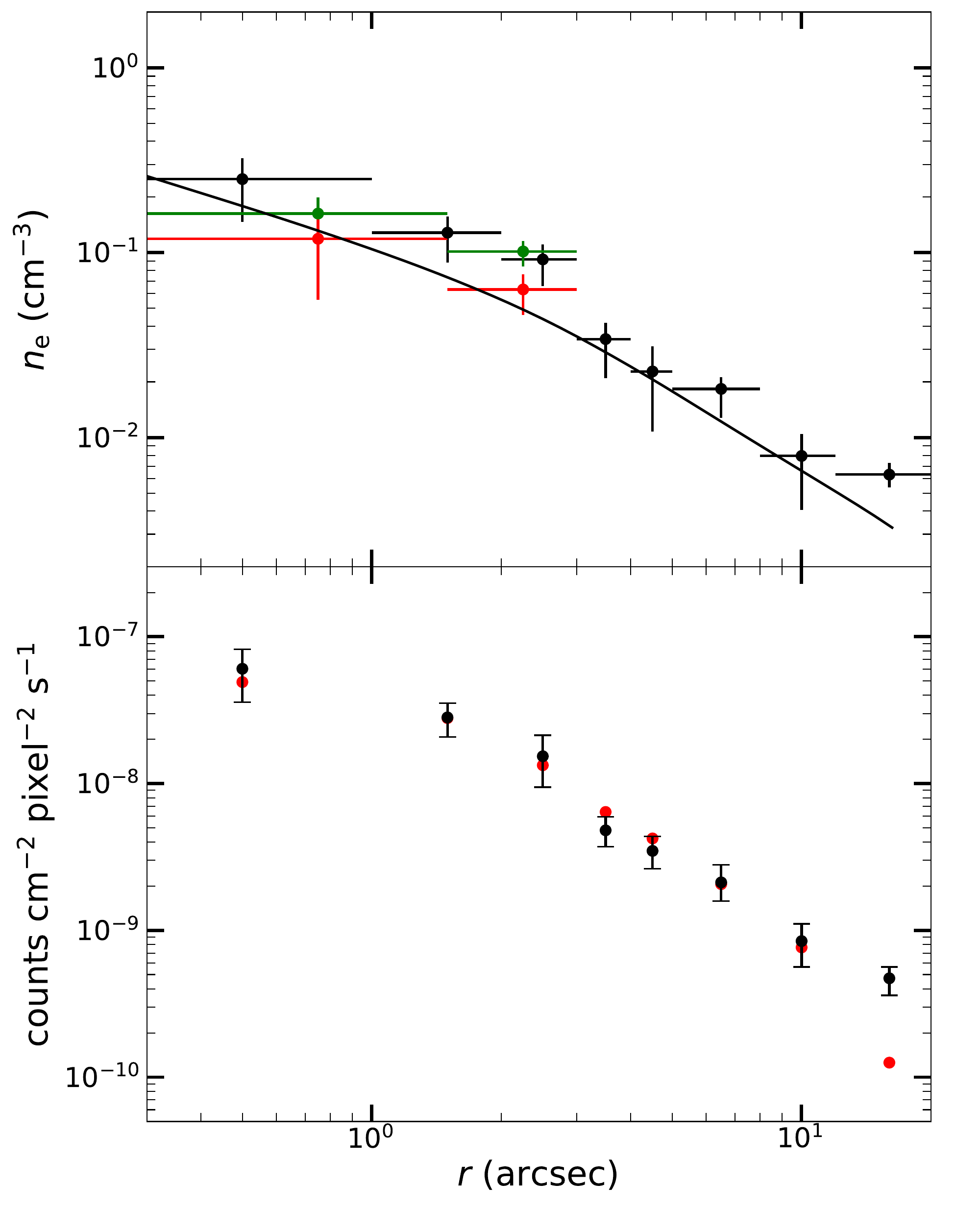}
\caption{The radial density and surface brightness profiles at 200~Myr compared with observations. In the top panel the solid line denotes the simulated density profile averaged over the whole angles and the black dots represent the deprojected density profile using the single temperature model in \citet{Wong:2014ko}. Red and green dots represent the hot and cool component, respectively, of the two-temperature model. Vertical error bars are at the 90\% confidence level and horizontal bars indicate the radial binning size. The bottom panel shows the projected surface brightness profiles of the simulations (red dots) and the observations for the hot gas (black dots with vertical error bars).
}
\label{fig:BestFit}
\end{figure}

Figure \ref{fig:BestFit} plots radial density and surface brightness profiles at 200~Myr compared with observations. In the top panel the solid line denotes the simulated density profile averaged over the whole angles and the black dots represent the deprojected density profile using the single temperature model in \citet{Wong:2014ko}. Red and green dots represent the hot and cool component, respectively, of the two-temperature model, which is preferred by the spectra fittings with respect to the single temperature model. In fact, \textit{Chandra} observations of the nearby galaxy M87 \citep{Russell:2015ec} also find the multi-temperature structure and the existence of H$\alpha$ filaments is supported by the observations \citep{2002ApJ...579..560Y,2004ApJ...607..294S}. Nevertheless, cooling is inefficient in our models because of the long cooling timescale. We suspect this discrepancy is caused by both our supernova feedback scheme that considers thermal energy injection alone and our inadequate numerical resolution. \cite{Li2015} adopt the momentum-driven supernova feedback model and find the multiphase medium is created easily by performing high resolution hydrodynamical simulations, in which the cooling radius of a supernova remnant is resolved by approximately 12~cells. Hot bubbles produced by supernovae can overlap and create the multiphase medium there above the critical supernova rate. Yet this momentum injection supernova feedback scheme of high resolution study is apparently beyond the scope of this paper, thus we just simply adopt the thermal injection method and focus on the fittings to the hot component, of which the density and temperature increase gently inward. In the bottom panel black dots with vertical error bars represent the observations of the surface brightness for the hot gas. Red dots denote our simulated surface brightness after conducting the projection.

As shown in Figure \ref{fig:BestFit}, the density and surface brightness profiles of our simulations agree well with the observations of the hot component. The deviations in the region outside 10\arcsec might be associated with the outer boundary condition. We adopt the outflow boundary condition and thus ignore the contribution from the stellar winds outside the simulation region. Yet winds with the circularisation radius smaller than the radius of the outer boundary would flow inward and increase the density in the vicinity of the outer boundary, thus causing the deviations there.

\begin{table*}
\centering
\begin{threeparttable}
\caption{Summary of the models.}
\label{tab:model}
   \begin{tabular}{lcccccc}
       \toprule
       Model & Viscosity & Stellar winds & Stellar potential& Supernovae  & Cooling & Thermal conduction \\
       \midrule
	   Run A   & $\surd $ & $\surd $   & $\times$     & $\times$     & $\times$  & $\times$ \\ 
	   Run B   & $\surd $ & $\surd $   & $\surd $     & $\times$     & $\times$  & $\times$ \\ 
	   Run C   & $\surd $ & $\surd $   & $\surd $     & $\surd $     & $\times$  & $\times$\\ 
	   Run D   & $\surd $ & $\surd $   & $\surd $     & $\surd $     & $\surd $   & $\times$\\ 
	   Run E   & $\surd $ & $\surd $   & $\surd $     & $\surd $     & $\surd $    & $\surd $\\ 
       \midrule
        	   Run Aa & $\times $ & $\surd $   & $\times$     & $\times$     & $\times$  & $\times$ \\ 
	   Run Ab & $0.1 $\tnote{a} & $\surd $   & $\times$     & $\times$     & $\times$  & $\times$ \\ 
	   Run Ac & $0.001$   & $\surd $   & $\times$     & $\times$     & $\times$  & $\times$ \\ 
       	   Run A1 & $\surd $ & 0.5\tnote{b}  & $\times$     & $\times$     & $\times$  & $\times$ \\ 
	   Run A2 & $\surd $ & 1   & $\times$     & $\times$     & $\times$  & $\times$ \\ 
	   Run A3 & $\surd $ & 1.5   & $\times$     & $\times$     & $\times$  & $\times$ \\ 
	   Run A4 & $\surd $ & 2   & $\times$     & $\times$     & $\times$  & $\times$ \\
	   Run B1 & $\surd $ & 0.5   & 0.5     & $\times$     & $\times$  & $\times$ \\ 
	   Run B2 & $\surd $ & 1  &  1     & $\times$     & $\times$  & $\times$ \\ 
	   Run B3 & $\surd $ & 1.5   & 1.5& $\times$     & $\times$  & $\times$ \\ 
	   Run B4 & $\surd $ & 2  & 2     & $\times$     & $\times$  & $\times$ \\  
	   Run B10 & $\surd $ & $\times$   & 0.5     & $\times$     & $\times$  & $\times$ \\ 
	   Run B20 & $\surd $ & $\times$  &  1     & $\times$     & $\times$  & $\times$ \\ 
	   Run B30 & $\surd $ & $\times$  & 1.5& $\times$     & $\times$  & $\times$ \\ 
	   Run B40 & $\surd $ & $\times$  & 2     & $\times$     & $\times$  & $\times$ \\  
       \bottomrule
   \end{tabular} 
\begin{tablenotes}
\footnotesize
\item[a] The numbers in the column of viscosity mean the viscosity coefficient $\alpha$.
\item[b] The numbers in the columns of stellar winds and potential represent the stellar density slope $q$, for which $\rho_*\propto r^{-q}$.
\end{tablenotes}
\end{threeparttable}
\end{table*}

The best radial density and surface brightness fittings are achieved at $f_\mathrm{q}=0.02$ and $v_\mathrm{w,sn}=350~\mathrm{km~s}^{-1}$. These two parameters are in charge of the density over the radius and the temperature at large radii, respectively. We suppose that our supernova injection scheme results in the small values of $f_\mathrm{q}$ and $v_\mathrm{w,sn}$. In fact, the energy contributions from supernova feedback should be larger than we adopt. Yet the temperature at large radii is sensitive to its value since we take the feedback in the form of internal energy. Hence, the supernova feedback would be stronger and produce more massive outflows if the momentum injection scheme is adopted \citep{Kim2015}. Correspondingly, $f_\mathrm{q}$ would be larger. In addition, \citet{1992ApJ...393..559K} report an ellipticity of $\epsilon \approx 0.4$ of the stellar mass distribution around the Bondi radius, implying shallower gravitational potential than that of the spherical assumption in our models. This would increase $f_\mathrm{q}$ further as discussed in \citet{Shcherbakov:2014cz}.


\subsection{Which process determines the density profile?}
\label{sec:components}

\begin{figure}
\centering
\includegraphics[width=0.5\textwidth]{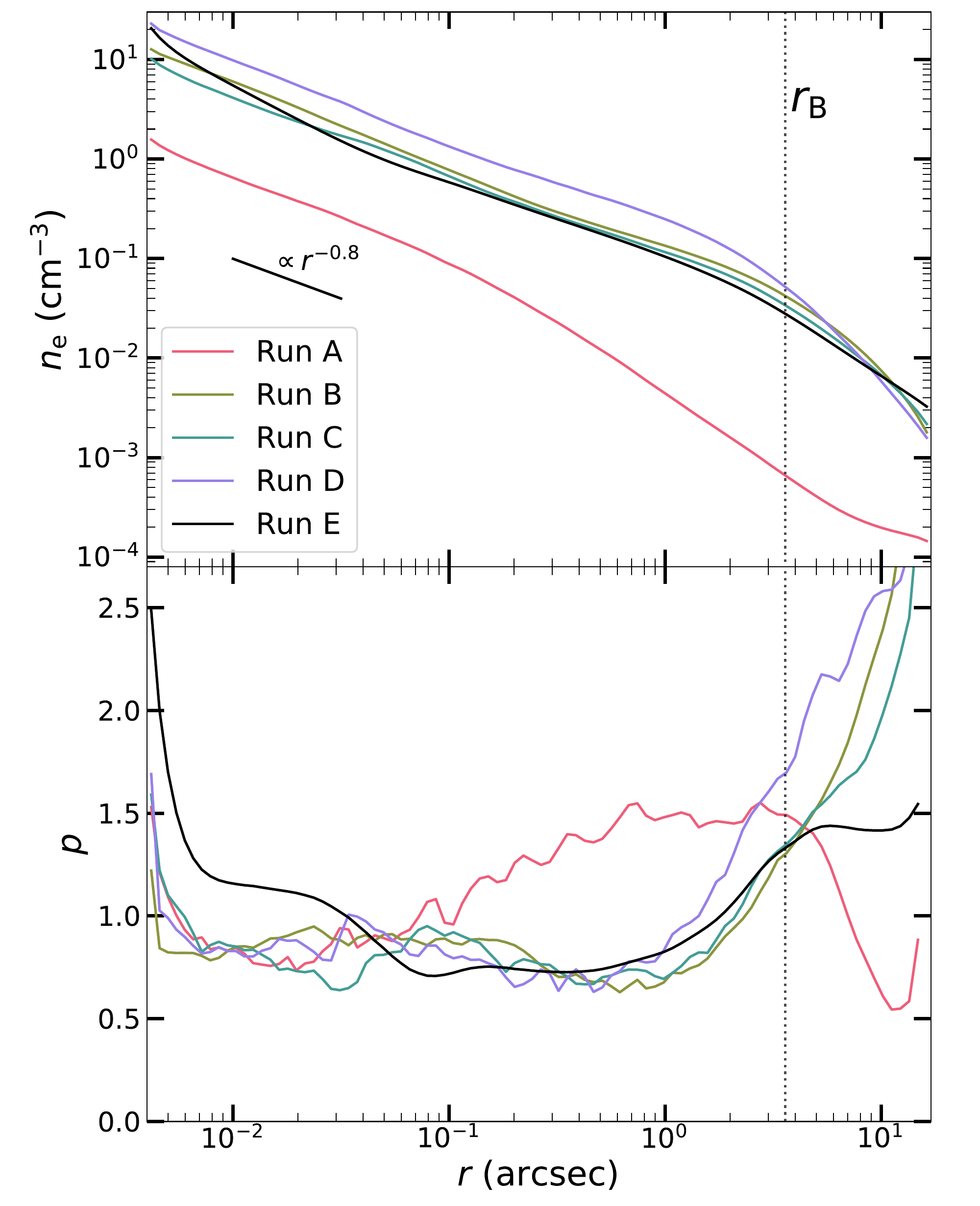}
\caption{The top panel shows the radial density profiles of various models averaged over angles and time. The bottom panel shows the density slope $p$ as a function of radius for each model. Models are listed in Table \ref{tab:model}. Particularly, Run~E is the best-fitting model.
}
\label{fig:Components}
\end{figure}

In this subsection, we aim to figure out the relative roles of various physical processes in determining the density profiles well inside and around the Bondi radius. We start by the simulations with stellar winds alone. Based on this simple case, we explore the role of viscosity in the accretion of stellar winds with moderate angular momentum. Then we add the stellar gravitational potential, supernova explosions, cooling, and thermal conduction in turn. The models are listed in Table~\ref{tab:model}. We average the density along the polar direction first concerning the snapshots between 190-200~Myr and then average over time to obtain smooth profiles. The radial density and slope profiles of all these models are shown in Figure \ref{fig:Components}. In Run~E, the simulation returns to the best fitting model mentioned above.

Run~A considers stellar winds alone. Stellar winds possess relatively high angular momentum at large cylindrical radii. The circularisation radius at the outer boundary of the midplane is larger than the Bondi radius. Yet the angular momentum becomes low at small cylindrical radii, suggesting that it decreases towards the poles at any radius. A similar ``funnel accretion'' scenario can be found inside the stagnation radius, which is $r_\mathrm{st}\sim0\farcs1$, more than an order of magnitude smaller than the Bondi radius. The density profile conforms to the RIAF theory $\rho \propto r^{-p}$ with $p=0.8$ inside the stagnation radius (\citet{Yuan2012a} find $p=0.85$ in the RIAF when $\alpha=0.01$).

\begin{figure}
\centering
\includegraphics[width=0.5\textwidth]{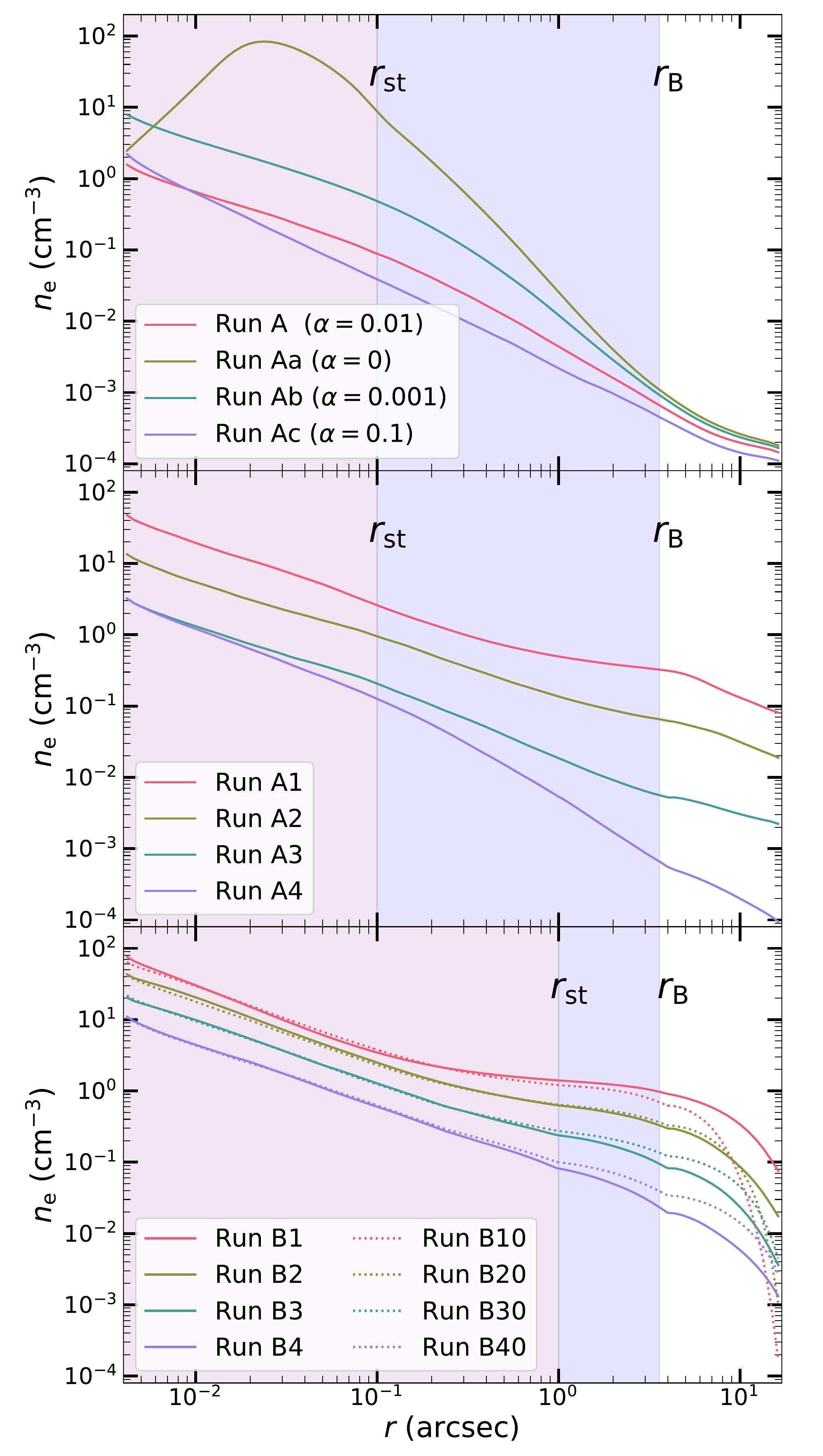}
\caption{The radial density profiles of various models listed in Table \ref{tab:model}. The top two panels show the models with the stellar winds alone, while the bottom panel shows the results of stellar gravitational potential with (solid line) or without (dotted line) stellar winds. In the bottom panel, the density profiles without stellar winds are normalised to make a better comparison with the profiles with stellar winds. The domain within the Bondi radius consists of two regions shaded by red and purple, respectively. They are bounded at the stagnation radius $r_\mathrm{st}$.
}
\label{fig:DenProfile}
\end{figure}

To provide some insight into the effect of viscosity on the radial density profile, we perform three simulations with different $\alpha$ on the basis of Run~A. The results are shown in the top panel of Figure~\ref{fig:DenProfile}. Run~Aa considers the stellar wind accretion without viscosity. Inside the stagnation radius, the accretion flow is much similar to the results in \citet{Proga2003}. Accretion dominates the polar region due to the low angular momentum, whereas the material that has too much angular momentum to be accreted forms a thick torus near the equator as shown by the peak of the olive green line. The thickness of the torus increases with time owing to the accumulation of the accreting materials until the inflow and outflow reach a balance. Consequently the accretion funnel becomes ``thinner'' and the mass accretion rate decreases. This is consistent with \citet{Proga2003} as they also find the mass accretion rate is constrained by the size and shape of the torus. Run~Ab explores the accretion process with a small viscosity coefficient $\alpha=0.001$ shown as the dark cyan line. The materials in the equator can be accreted with the assistance of the angular momentum transfer, thus unlike Run~Aa, the density increases with the decreasing radius. The density is still higher than that in Run~A due to less efficient angular momentum transport, but the slope is similar inside the stagnation radius. By contrast, Run~Ac considers a large viscosity coefficient $\alpha=0.1$ shown as the purple line. The density is generally lower than that of Run~A since the accretion is more violent when the angular momentum transport is more efficient. Indeed, we find that the net accretion rate at the inner boundary is an order of magnitude higher than that of Run~A. The radial density profile tends to be steeper in the inner region with $p=1.2$, caused by the weaker convective instability when the stress is stronger. This is also found in many other simulations \citep{Igumenshchev1999,Stone1999,Yuan2012a}.
The density of all these four simulations with different $\alpha$ converge at large radii, where the gas dynamics is dominated by the outflow. Therefore, viscosity only has an effect during the accretion process in the inner region.

The density profile of Run~A becomes steep outside the stagnation radius, suggesting the potential role of the stellar winds in shaping the radial density profile. To further investigate the role of stellar winds, we perform a set of simulations with the same setups as Run~A but different stellar density slopes listed in Table \ref{tab:model}. The results are shown in the middle panel of Figure \ref{fig:DenProfile}. The stellar density is calibrated with the same enclosed stellar mass at the Bondi radius, where the enclosed stellar mass should be comparable to the BH mass. Within the Bondi radius, the domain could be divided into two parts by density slopes bounded at the stagnation radius. Inside this radius, which is shaded by red, Run~A1-A3 show the same density slope $p=0.8$, yet Run~A4 possesses steeper density profile. This is because the stellar winds injection from the inner shells $\propto 4\pi r^2\rho_*$ are equal to those from the outer shells if $\rho_* \propto r^{-2}$. Consequently the injection timescale of the stellar winds is shorter than the viscous timescale and gas density profile tends to approximate to the stellar density distribution. Between the stagnation radius and the Bondi radius, which is shaded by purple, the theory of the RIAF may be not applicable, since the gas flow is mainly outflow. The gas density profile in this region is strongly affected by the stellar density slope. With increasing stellar density slope $q$, the density profile of the gas  steepens. 

Run~B considers the contribution from stellar gravitational potential. With combined gravity, the potential well turns deeper and gas density grows larger with a break radius around the Bondi radius. Correspondingly, the stagnation radius $r_\mathrm{st}\sim1\arcsec$ becomes larger as well. Compared with Run~A, the density profile inside the stagnation radius remains flat with $p=0.8$, yet the profile between $0\farcs1 - 1\arcsec$ becomes shallower. This can be explained by the RIAF in the combined gravitational potential with shorter viscous timescale than the injection timescale. Around the Bondi radius, the profile transitions from a steep profile with $p>2$ outside to $p=0.8$ inside the Bondi radius smoothly. This significantly differs from the case without stellar potential in Run~A, of which the profile transitions from $p=0.5$ outside to $p=1.5$ across the Bondi radius. \citet{Wong:2014ko} also find the steepening of the density profile outside the Bondi radius after considering the galactic potential on the Bondi flow (their Figure~8), which is consistent with our results.

We also perform simulations Run~B1-B4 to explore the consequences under different potential profiles, and the results are presented as solid lines in the bottom panel of Figure \ref{fig:DenProfile}. The stagnation radius of all the models is around $1\arcsec$ in spite of the minor differences. In the red-shaded region the density slopes of the models with $q>1$ are almost the same as Run~B. It is interesting to find that with stellar potential the density profile of Run~B4 inside $0\farcs1$ becomes shallower compared with Run~A4. This may be related to the shorter viscous timescale than the injection timescale with the assistance of the large enclosed stellar mass. In the purple-shaded region the profiles of all the models generally become shallower. A simple explanation is given as follows. Given the minor role of cooling, the adiabatic gas flow produces $\nabla P/\rho \propto r^{-p(\gamma-1)-1}$ with $\rho\propto r^{-p}$. The gravity from stars is $\mathrm{G}M_\mathrm{enc}/r^2 \propto r^{-q+1}$ with $\rho_*\propto r^{-q}$ and the gravity from the BH is $\mathrm{G}M_\mathrm{BH}/r^2 \propto r^{-2}$. The pressure gradient $\nabla P/\rho$ is proportional to the combined gravity $\mathrm{G}(M_\mathrm{enc}+M_\mathrm{BH})/r^2$ in the steady state since the centrifugal force is a constant fraction of the gravity in the RIAF. Hence in the region where stellar potential works, the inclusion of the stellar gravity yields a shallower gravity profile as long as $q<3$. Moreover, a larger $q$ results in a larger $p$ to balance the gravity, which is consistent with the one-dimensional calculations in \citet{Quataert:zz}. Outside the Bondi radius, all these four models show the steepening of the density profiles. Therefore, compared with stellar winds, the stellar potential has a more significant influence on the density profile, especially in the vicinity of the Bondi radius. Beyond the Bondi radius a smooth transition can be found in the density profiles of all the models.

To verify the dominant role of stellar potential, Run~B10-B40 ignore the contribution of stellar winds on the basis of Run~B1-B4, respectively. The density profiles are normalised to fit the profiles of Run~B1-B4 in the red-shaded area. By comparing the density profiles of Run~B10-B40 with those of Run~B1-B4, we find the stellar winds tend to steepen the profiles in the purple-shaded region except in Run~B1, yet the differences between them are trivial, implying the dominant role of stellar potential in shaping the radial density profiles of the hot gas flow. 

Run~C adds supernovae into the simulation. With little variation in the radial density profile, supernova feedback raises the temperature at large radii where it becomes more important based on the timescale analysis.

Run~D incorporates cooling in the simulation. The temperature decreases gently as expected, while the density increases correspondingly. Yet the density profile inside the Bondi radius varies little, which is consistent with the inefficiency of cooling in the RIAF.

In the end, Run~E includes thermal conduction and the simulation returns back to our best-fitting model. Because the energy flux is transported outward though thermal conduction, the temperature profile turns shallower and the gas density becomes lower correspondingly due to the stronger mass outflow, thus resulting in the lower mass accretion rate into the centre. The result is consistent with the one-dimensional calculations in \citet{Shcherbakov:2014cz} that find the density in advective models is larger than that in conductive models. With regard to the density profile inside the Bondi radius, the profile steepens inside 0\farcs1. This is possibly unphysical because the unsaturated thermal conduction produces much stronger heat flux than it should be under saturation. Beyond the Bondi radius, the profile becomes shallower due to the relative importance of thermal conduction at large radii as indicated by Figure~\ref{fig:timescale}.

In a summary, we find that stellar potential is the dominant process in determining the density profile around the Bondi radius, where the profile transitions smoothly from the slope $p>1$ to $p<1$. Viscosity has an effect in the central region, while thermal conduction plays a role in flattening the profile outside the Bondi radius. Note that we do not focus on the difference of the density slope at exactly the Bondi radius, because it has large uncertainties considering the range of the Bondi radius and the parameters we choose due to the resolution of the observations.

\subsection{Comparisons with one-dimensional results}\label{sec:Comp1D}
\citet{Shcherbakov:2014cz} have conducted one-dimensional calculations on the same source with the similar model. The major difference between our works is the treatment of angular momentum transport. We take advantage of the additional dimension to account for the angular momentum transfer in our simulations, whereas they simply treat the angular momentum transport as a posteriori. However, a general consistency is found between our results. Averaging over the whole angles, our simulations show a characteristic radius where the averaged radial velocity is zero. This radius is quite similar to the stagnation radius in one-dimensional calculations, and they both are of the value $r_\mathrm{st}\sim0.1~r_\mathrm{B}$. The angle-averaged density and temperature agree with one-dimensional calculations as well, since we aim to fit the observational data.

In spite of these similarities, the angular profiles of these properties in two-dimensional simulations reveal a totally different accretion pattern. Within the stagnation radius, winds are mainly accreted in the polar region due to the low angular momentum there, whereas materials with high angular momentum in the midplane become outflow or convection, since the angular momentum of the stellar winds at a given radius decreases towards the pole. This ``funnel'' accretion scenario is also found in other simulations with spherical-like distribution of rotationally accreting material \citep{Proga2003,2018arXiv180500474R}. The density and temperature at the equatorial plane are relatively higher than those in the polar region, thus producing the highest mass accretion rate at the boundary of polar and disk regions. Therefore the accretion pattern of the two-dimensional simulations is totally different from the inflow-outflow structure in the one-dimensional calculations. Another difference between our results is that the parameters of our best-fitting model $f_\mathrm{q}$ and $v_\mathrm{w,sn}$ are generally smaller than those in one-dimensional calculations. The smaller $f_\mathrm{q}$ can be explained by three reasons. First, we adopt twice higher BH mass, which results in deeper gravitational potential that allows for more massive gas reservoir. Thus $f_\mathrm{q}$ should be smaller to achieve the observational gas density. Second, our outer boundary is set at $17\arcsec$, which is larger than the radius of one-dimensional outer boundary $12\arcsec$. The larger outer boundary causes the mass injection rate by stellar winds to increase by 50 per cent. Since stellar winds are the only mass source in our models, a smaller $f_\mathrm{q}$ is required to cancel out this extra mass injection. Third, a smaller $v_\mathrm{w,sn}$ could also lead to a smaller $f_\mathrm{q}$ due to less efficient heating by supernova explosions. The smaller $v_\mathrm{w,sn}$ is constrained by the temperature at large radii, especially the temperature between 10\arcsec and 20\arcsec, which is not fitted in the one-dimensional calculations.

\section{Conclusions}\label{sec:conclusions}
In this paper we perform simulations with various physical processes on the hot gas flow around the Bondi radius in the LLAGN NGC\,3115, extending the previous one-dimensional calculations to two-dimension. Along the same lines as one-dimensional models, we obtain the properties of the nuclear star cluster from observations. We consider the mass, energy, and angular momentum injection from stellar winds and the energy contribution from supernova explosions. The physical processes are incorporated including gravitational pull by the enclosed stellar mass, radiative cooling, and thermal conduction. The best fittings of density and temperature are achieved at $f_\mathrm{q}=0.02$ and $v_\mathrm{w,sn}=350~\mathrm{km~s}^{-1}$, which control the mass loss rate of stars and energy injection of Type Ia supernovae, respectively, and agree generally well with the observations of \textit{Chandra}.

Observations suggest that the density slope $p$ around the Bondi radius is $p\sim1$, which deviates from the simulation results about the RIAF. Our simulations find that the theory of RIAF applies inside the stagnation radius $r_\mathrm{st}\sim0.1~r_\mathrm{B}$ and $p\sim0.8$. This radius satisfies that the angle-averaged radial velocity equals zero. Outside this radius outflow becomes significant. The density profile tends to be steep as the radius goes outward, producing the density slope $p\sim1$ around the Bondi radius. After exploring the role of each physical process by performing a set of simulations, we find that it is stellar potential that mainly determines the density slope around the Bondi radius. In addition, thermal conduction also plays a role in flattening the profile outside the Bondi radius.



Our simulation models are based on the one-dimensional work by \citet{Shcherbakov:2014cz}. After averaging the whole angles, the properties of the hot gas, such as density, temperature, and radial velocity, are quite similar to their one-dimensional results. Considering angular momentum, however, their angular profiles exhibit the distinctions and features of two-dimensional simulations inside the stagnation radius. We find the gas in the polar regions is accreted directly without the assistance of angular momentum transfer, and flows outward within the thick disk, of which the opening angle is $\sim120\degr$. It is worth noting that the mass accretion rate at the inner boundary is dominated by the accretion at the surface the disk, and viscosity still plays an important role in the accretion process.


At the end of this paper, we acknowledge that there are several limitations in our work that may affect the fittings to the observations and could be improved in the future. First, we assume zero ellipticity $\epsilon=0$ of the stellar distribution for simplicity, which results in deeper gravitational potential than that of $\epsilon\sim0.4$ suggested by \citet{1992ApJ...393..559K}. Deeper gravitational potential leads to higher density and temperature of the gas flow. Therefore, both $f_\mathrm{q}$ and $v_\mathrm{W,sn}$ should be larger after considering a non-zero ellipticity. Second, we do not consider inflow from the outer boundary, which may underestimate the density around the outer boundary. Third, we inject supernova feedback as a pure thermal energy injection, which might result in over-heating and less turbulence. Both $f_\mathrm{q}$ and $v_\mathrm{W,sn}$ should be larger as well if a more realistic supernova feedback scheme is adopted. In this work we do not focus on the influences of supernovae on the hot gas flow around the Bondi radius, and the conclusions on the radial density profile do not vary qualitatively with these caveats. It is interesting to use three-dimensional simulations with inflow and a realistic supernova feedback scheme to study the growth of thermal instability and formation of cold filaments in the context of the nearby galactic nucleus, so that simulation results can be compared to observations directly.

\section*{Acknowledgements}\label{sec:acknowledgements}
We thank the anonymous referee for his/her useful suggestions and Ka-Wah Wong for his observational data. We also thank Feng Yuan, Sean Ressler, Bocheng Zhu, Weixiao Wang, Junjie Feng for useful discussions. This work is supported in part by the National Key Research and Development Program of China (Grant No. 2016YFA0400704), the Natural Science Foundation of China (grants 11573051, 11633006, 11650110427, 11661161012), and the Key Research Program of Frontier Sciences of CAS (No. QYZDJSSW-SYS008). Z.G. was supported in part by the Natural Science Foundation of Shanghai (grant 18ZR1447200). This work made use of the High Performance Computing Resource in the Core Facility for Advanced Research Computing at Shanghai Astronomical Observatory.




\bibliographystyle{mnras}
\bibliography{NGC3115}







%
\bsp	
\label{lastpage}
\end{document}